\documentclass[sigconf]{acmart}
\usepackage{tabularx}
\usepackage{makecell}
\usepackage{threeparttable}
\usepackage[utf8]{inputenc} 

\usepackage{booktabs}       
\usepackage{nicefrac}       
\usepackage{microtype}      
\usepackage{xcolor}         
\usepackage{soul}
\usepackage{url}
\usepackage[font={small,bf}]{caption}
\usepackage{graphicx}
\usepackage{amsmath}
\usepackage{amsthm}
\usepackage{algorithm}
\usepackage{algorithmic}
\usepackage{multirow}
\usepackage{subfigure}
\usepackage{bm}
\usepackage{color}

\usepackage{tablefootnote}
\usepackage{lipsum}
\usepackage{enumitem}

\usepackage{xspace}

\usepackage{stfloats}
\usepackage{marvosym}

\usepackage{pifont}
\newcommand{\ie}{\emph{i.e.}}
\newcommand{\eg}{\emph{e.g.}}

\newcommand{\R}{\mathbb{R}}
\newcommand{\vect}[1]{\mathbf{#1}}
\newcommand{\mat}[1]{\mathbf{#1}}
\newcommand{\Attn}{\operatorname{Attn}}
\newcommand{\PEpos}{\operatorname{PE}_{\text{pos}}}
\newcommand{\PEdel}{\operatorname{PE}_{\Delta}}
\newcommand{\softmax}{\operatorname{softmax}}

\setcopyright{none}

\usepackage{color, colortbl}

\begin{document}

\title{Practice on Long Behavior Sequence Modeling \\ in Tencent Advertising}

\author{Xian Hu$^{1*}$, Ming Yue$^{1*}$, Zhixiang Feng$^{1*}$, Junwei Pan$^{1\dagger}$, Junjie Zhai$^{1}$, Ximei Wang$^1$ \\
Xinrui Miao$^2$, Qian Li$^1$, Xun Liu$^1$, Shangyu Zhang$^1$, Letian Wang$^1$, Hua Lu$^1$, Zijian Zeng$^1$ \\
Chen Cai$^1$, Wei Wang$^1$, Fei Xiong$^1$, Pengfei Xiong$^1$, Jintao Zhang$^1$, Zhiyuan Wu$^1$
\\ Chunhui Zhang$^1$, Anan Liu$^1$, Jiulong You$^1$, Chao Deng$^1$ \\
Yuekui Yang$^1$, Shudong Huang$^1$, Dapeng Liu$^1$, Haijie Gu$^1$
}
\affiliation{
$^1$Tencent Inc.\country{}
$^2$Independent\country{}
$^*$equal contribution
$^\dagger$corresponding author
}

\email{{sirlyhu, rominyue, lionelfeng, jonaspan, jasonzhai}@tencent.com}

\renewcommand{\shortauthors}{Junwei et.al.}

\begin{abstract}

    Long-sequence modeling has become an indispensable frontier in recommendation systems for capturing users' long-term preferences. 
    However, user behaviors within advertising domains are inherently sparse, posing a significant barrier to constructing long behavioral sequences using data from a single advertising domain alone. 
    This motivates us to collect users' behaviors not only across diverse advertising scenarios, but also beyond the boundaries of the advertising domain into content domains—thereby constructing \emph{unified commercial behavior trajectories}.
    This cross-domain/scenario integration gives rise to the following challenges: 
    (1) \emph{feature taxonomy gaps} between distinct scenarios and domains, 
    (2) \emph{inter-field interference} arising from irrelevant feature field pairs, 
    and (3) \emph{target-wise interference} in temporal and semantic patterns when optimizing for different advertising targets. 
    To address these challenges, we propose several practical approaches within the two-stage framework for long-sequence modeling.
    In the first (search) stage, we design a hierarchical hard search method for handling complex feature taxonomy hierarchies, alongside a decoupled embedding-based soft search to alleviate conflicts between attention mechanisms and feature representation. 
    In the second (sequence modeling) stage, we introduce: 
    (a) Decoupled Side Information Temporal Interest Networks (TIN) to mitigate inter-field conflicts; 
    (b) Target-Decoupled Positional Encoding and Target-Decoupled SASRec to address target-wise interference; 
    and (c) Stacked TIN to model high-order behavioral correlations.
    Deployed in production on Tencent’s large-scale advertising platforms, our innovations delivered significant performance gains: an overall 4.22\% GMV lift in WeChat Channels and an overall 1.96\% GMV increase in WeChat Moments.
   
\end{abstract}

\maketitle

\section{Introduction}

Online advertising has become a billion-dollar business nowadays, with an annual revenue of 225 billion US dollars between 2022 and 2023 (increasing 7.3\% YoY)~\cite{iab-report}. 
One of the most critical tasks is to predict the Click-Through Rate (CTR) for target items based on a user's behavior sequence.
Previous research has demonstrated that longer user histories contribute to more precise predictions~\cite{piqi2020sim}. 
Consequently, there has been a growing interest in long-sequence user interest models in recent years~\cite{chenqiwei2021ETA, cao2022sampling, chang2023TWIN, si2024TWINV2, hou2022towards, feng2024DARE}.

To satisfy the real-time inference latency requirement, existing long-sequence models usually employ a two-phase process~\cite{piqi2020sim}: \textit{search} (also known as the General Search Unit) and \textit{sequence modeling} (also known as the Exact Search Unit). 
The search phase aims to retain the most informative behaviors from the long sequence via hard search on a specific feature, such as category~\cite{piqi2020sim} or soft search by attention weights~\cite{piqi2020sim, chang2023TWIN}, while the sequence modeling aims to extract the user's interest upon the retained behaviors from the search stage~\cite{zhouguorui2018DIN, zhouguorui2019DIEN, fengyufei2019DSIN, chenqiwei2021ETA, zhouhaolin2024TIN}.

In advertising recommendations, users' behaviors regarding ads are much sparser than those on content. 
This is primarily due to the user's reluctance to view, not to mention providing positive feedback such as clicking or converting on the ads.
Besides, Tencent's platform has strict restrictions on the frequency and total number of ad exposures for a better user experience.
To build a long behavior sequence to capture a user's long-term interest effectively, besides users' ad behaviors of the current traffic scenario, we also collect available behaviors not only from other ad scenarios but also, crucially, the content domain. 
The user's abundant behaviors on the content domain contain more collaborative signal, which reflects her interest.
However, feature taxonomies vary across different ad traffic scenarios, and the disparity between advertisement and content domains is particularly pronounced, posing significant challenges for long-sequence modeling:

\begin{itemize}
    \item \textbf{C1: Inter-field Interference.} First, these heterogeneous fields from different advertising scenarios and the content domain lead to severe inter-field interference within the attention since they introduce many noisy interactions~\cite{tupe2020rethinking}, such as \texttt{(behavior position) X (target category)}.
    \item \textbf{C2: Target-wise Interference.} Second, the cross-scenario and cross-domain trajectories make the semantic and temporal correlations entangle with each other, especially for different target categories, leading to interference between advertising targets. For example, a user's interest in her persistent hobby, such as Hiking, may have a less decayed temporal pattern than in Breaking News. Using a global temporal encoding may fail to capture such diverse patterns.
    \item \textbf{C3: High-order Correlation.} Third, with so many fields, there should be high-order correlation between them, such as \texttt{(Behavior Category) X (Target Category) X (Behavior Scenario) X (Target Scenario)}. Most existing sequential ranking models, \emph{i.e.}, target attention methods~\cite{zhouguorui2018DIN, fengyufei2019DSIN, chenqiwei2021ETA, zhouhaolin2024TIN}, only employ one target-attention layer, failing to capture such complex correlation.
\end{itemize}

In this paper, we present solutions to address the aforementioned and other related challenges.
Specifically, in the search stage, we first propose a hierarchical hard search policy for ranking, which effectively handles the heterogeneous side features of the sequence (C1). 
We also introduce the DARE method~\cite{feng2024DARE} for soft search, which decouples attention and representation embedding to resolve the conflicts between these two modules (C1).
For the matching where the target is not available, we employ a stratified sampling method that samples behaviors from each category to preserve the user's comprehensive interest.

Next, in the sequence modeling phase, we adopt the Temporal Interest Network (TIN) to capture both the semantic and temporal correlation of behavior sequences.
We first present the decoupled side info TIN to handle the inter-field interference by employing multiple TINs and decoupling feature fields in the attention modules of them (C1).
We then demonstrate target decoupled position encoding and target decoupled SASRec to tackle target-wise interference by decoupling the position encoding and the output heads (C2).
At last, we present stacked TIN to model high-order correlation between features, with a novel stacking design of the target-attention module (C3). 

We also present some model training and inference improvements to support long-sequence modeling, including a parameter prefetch strategy for efficient exchange, a data prefetch module for GPU acceleration, GPU Key-Collection Acceleration, and Multi-Stream designs to improve GPU utilization, and a set of universal heterogeneous acceleration engines for efficiency.

The rest of the paper is organized as follows: Sec.~\ref{sec:trajectory} describes how to construct long sequences for advertising, Sec.~\ref{sec:method} presents our proposed approaches to tackle the challenge of inter-field and target-wise feature interference, as well as high-order feature interaction, Sec.~\ref{sec:platform} demonstrates our platform improvement techniques.

\section{Construct Unified Commercial Behavior Trajectories}
\label{sec:trajectory}

\begin{figure}
    \centering
    \includegraphics[width=0.7\linewidth]{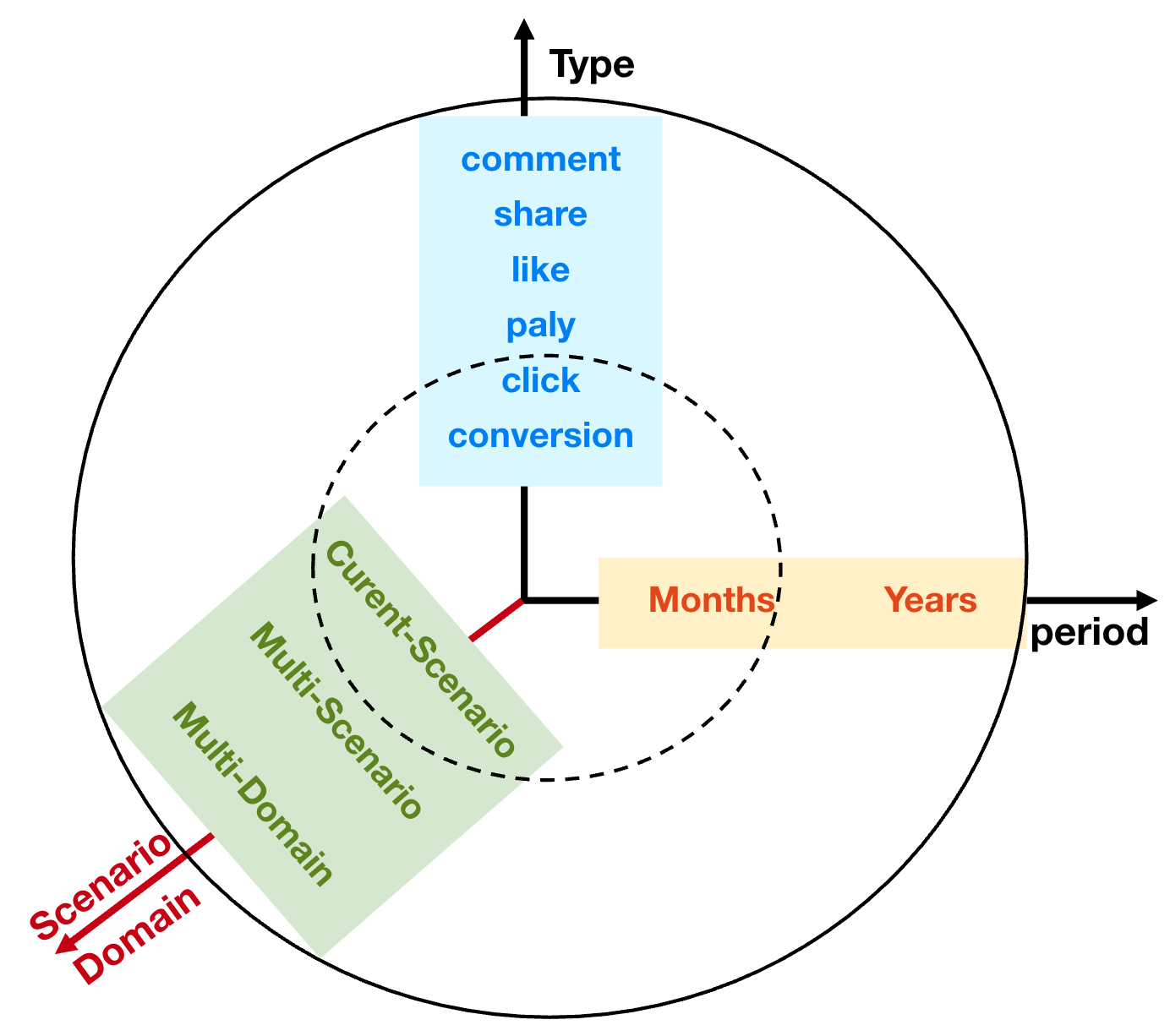}
    \caption{Illustration of building up a long sequence for advertisement recommendation.}
    \label{fig:features}
\end{figure}

Different from e-commerce recommendations~\cite{piqi2020sim, hpmn2019, chenqiwei2021ETA, cao2022sampling} or content recommendations~\cite{chang2023TWIN, si2024TWINV2, hou2022towards}, where there are plenty of user behaviors, a user's behaviors on the advertisement are much sparse, making it hard to collect enough behaviors to capture her comprehensive and long-term interest.
To this end, in addition to simply including all behaviors from a given advertising scenario, we also need to involve behaviors of various types from other scenarios or even from the content recommendation domain.
Specifically, we focus on the following two aspects when building up the long sequence for ads recommendation:

\begin{table*}[h!]
\begin{tabularx}{\linewidth}{p{4cm}XX}
\toprule
\textbf{Image:} & \includegraphics[height=2cm]{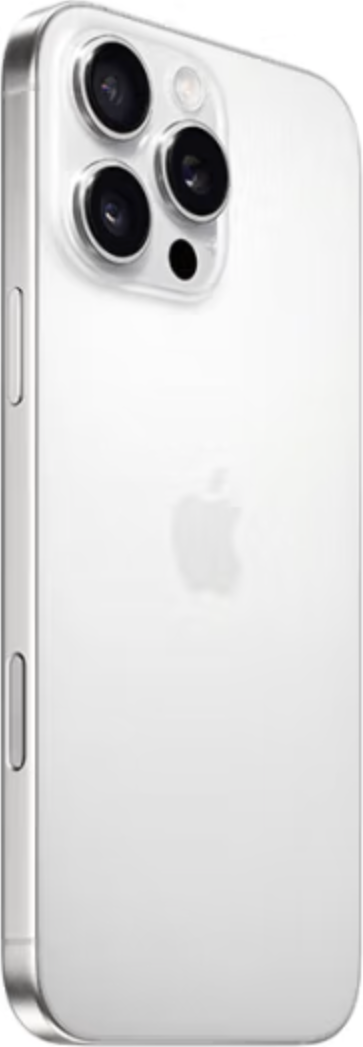} & \includegraphics[height=2cm]{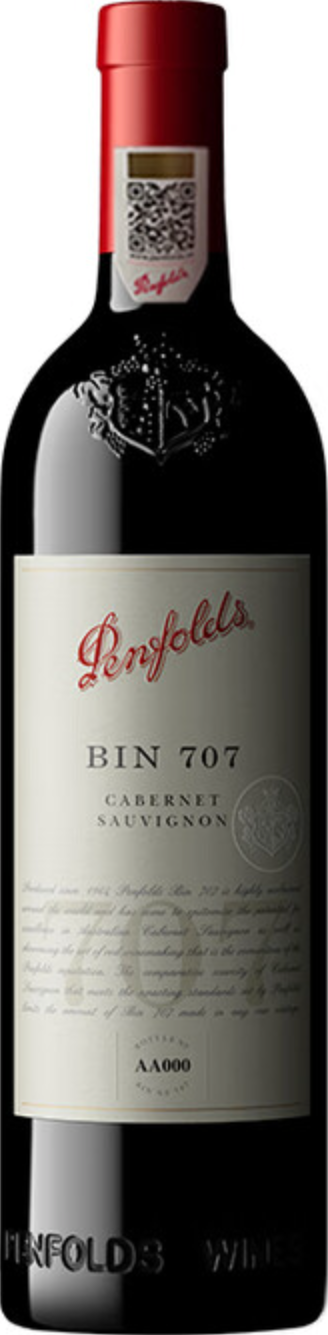}\\
\midrule
\textbf{Title:} & Apple iPhone 16 Pro Max (A3297) 256GB White Titanium Supports China Mobile/Unicom/Telecom 5G Dual SIM Support & Penfolds Cabernet Sauvignon Red Wine 750ml (Bin707) Australian Imported Genuine Official Product Valentine's Day Gift \\
\midrule
\textbf{Step 1: Identify Category} & Mobile Communications → Mobile Phones & Alcohol → Wine → Red Wine \\
\midrule
\textbf{Step 2: Retrieve Properties} & Brand, Model & Brand, Origin, Version, Sweetness, Net Content \\
\midrule
\textbf{Step 3: Identify Values} & \makecell[tl]{Brand: Apple \\ Model: iPhone 16 Pro Max} & \makecell[tl]{Brand: Penfolds \\ Origin: Australia \\ Version: Bin707 \\ Sweetness: Dry \\ Net Content: 750ml} \\
\midrule
\textbf{Step 4: Create SPU} & Apple iPhone 16 Pro Max Mobile Phone & Penfolds Australia Dry 750ml Red Wine \\
\bottomrule
\end{tabularx}
\caption{Two Cases of SPU Construction.}
\label{tab:SPU_construction}
\end{table*}

\paragraph{Behavior Consolidation}
Existing long sequence modeling~\cite{piqi2020sim, chenqiwei2021ETA,  chang2023TWIN, si2024TWINV2, hou2024cross} in the industry primarily focus on information flow or e-commerce scenarios, where the items in user behavior sequences are from the same scenario, same domain, and same action type as the recommended items. 
However, in Tencent Advertising platform, the behavioral volume of individual scenarios is limited, making it challenging to achieve significant improvements. 
Therefore, we aim to break through the constraints of the same type, same domain, and same scenario to consolidate all available user behaviors:

\begin{itemize}[leftmargin=*]
    \item \textbf{Type expansion}: In addition to user clicks and conversion behaviors, we include additional behaviors such as video play information, likes, follows and comments etc.
    \item \textbf{Scenario expansion}: User's ads behavior may spread across multiple scenarios such as WeChat Moments, Channels or Official Accounts, even Tencent News or Video. We include user's behaviors from all ads scenarios to represent her ads interest.
    \item \textbf{Domain expansion}: User behavior on the content domain is hundreds of times more than ad domain, and content behavior with commercial intention better reflects users' comprehensive interests.    
\end{itemize}

\paragraph{Longer Length and Period}
We collect more user behaviors by extending the period as well as the length.
Recognizing the significance of advertising domain data, we extend the data from months to years to capture users' long-term and stable interests.
By combining content domain data, we construct sequences with tens of thousands of items, accommodating a greater number of user behaviors and significantly enriching the average number of user interests (from dozens to thousands).

Including user's behaviors of various types from different scenarios or domains bring several unique challenges:

\begin{itemize}[leftmargin=*]
    \item \textbf{Noisy Behavior Types}: Non-interactive behaviors such as ad playback lack clear user intention and may overwhelm strongly correlated behaviors with fewer occurrences
    \item \textbf{Different Behavior Representation}: Behaviors are collected across various scenarios or domains, with possibly very different feature sets. This feature gap makes it hard to share knowledge across these behaviors.
    \item \textbf{Feature Polysemy}: Behaviors such as playback and reading vary across scenarios or domains, necessitating careful behavior and item representations.
\end{itemize}

\paragraph{Our Solution}

In theory, it is ideal to retain complete user behavior, including actions, behavior domains (items), and other behavior-related information. 
However, preserving and learning from raw information incurs significant resource and exploration costs. 
Therefore, during the feature engineering stage, we attempt to refine the original user behavior data, include more features on the behaviors, such as behavior scenario, playtime, and conversion type, and expand different granularity dimensions of side information based on behavior domains (items).

First, we construct \emph{a unified commercial behavior trajectory} by combining use's behaviors across types, scenarios, and domains with the shared features such as the Standard Product Unit (SPU) and product categories.
\emph{The product category feature plays a critical role in the models} (especially in the hard search, stratified sampling and target decoupled SASRec for Matching) as a unified feature across all behaviors.
Refer to Sec.~\ref{subsec:spu} for details.

Second, we include all side info representing each behavior's type, source scenario, or domain to decouple them. 
With all these side info and our model's ability to decouple them(details in Sec.~\ref{subsec:sequence_modeling}), we succeed in handling these heterogeneous behaviors.

Third, we also include some fine-grained side information for each behavior, including creative fingerprints, product fingerprints, advertiser accounts, and fine-grained level categories, \emph{e.g.}, the 2nd to 3rd-level categories. 
The combination of these fine-trained features with the coarse-grained ones, such as SPU or category of various levels, facilitates the hierarchical hard search policy as well as the sequence modeling.

Finally, our long behavior sequences contain tens of thousands of behaviors, each behavior comprising detailed information on $(user,\, item,\, action\ time,\, action\ specifics)$. The side information for each item covers ten aspects: item id, creative fingerprint, product fingerprint, SPU, product category levels 1 to 3, marketing target, advertiser account, and ad industry. The action specifics are described by three attributes: action type, action scenario, and action detail.

\subsection{SPU Creation and Identification}
\label{subsec:spu}

\paragraph{Definition of SPU}

SPU (Standard Product Unit) is the fundamental unit for product information management.
It aggregates attributes related to the essence of the product (such as brand, model, material) to form a reusable, easily searchable minimal information unit.
It establishes a structured representation of identical products at the finest granularity.

SPU is usually extracted from the CPV (Category-Property-Value) system, which includes:
\begin{itemize}
    \item \textbf{Category}: Categories are typically a multi-level tree structure used for hierarchical classification to locate products. 
    E.g.: Mobile Communications → Mobile Phones, Alcohol → Wine → Red Wine.
    \item \textbf{Property and Value}: Based on the category, properties and values further characterize the specific features and selling points of a product, presented as multiple \texttt{<key:value>} pairs.
    E.g.: \texttt{<Efficacy: Whitening>}, \texttt{<Material: Cotton>}.
\end{itemize}

Our process for creating an SPU is as follows:

\textbf{Step 1}: Offline, a comprehensive SPU library is established by processing massive product data through stages including quality cleansing, CPV understanding based on VLM (Visual Language Model), clustering, and deduplication. The goal is to cover all products in the domain without duplication.

\textbf{Step 2}: Each newly created advertisement is bound to a product. Each product is then matched to the most relevant SPU in the library via a relevance algorithm, ultimately building the mapping relationship: \texttt{Advertisement → Product → SPU}. The \texttt{Product → SPU} process is illustrated in Tab.~\ref{tab:SPU_construction}.

As a result, we construct an SPU pool with 700,000  SPU IDs, with a precision of 97\% and a duplication rate of 2.73\%. 
The overall mapping accuracy for \texttt{Product → SPU} process is more than 90\%.

\section{Method}\label{sec:method}

The heterogeneity of behaviors of various side information in the above-mentioned unified commercial trajectory brings huge challenges to both search and sequence modeling.
We'll present our solutions in this section.

\paragraph{Problem setup and notation.}
A user's unified commercial trajectory is a sequence
$\mathcal H=(x_1,\ldots,x_n)$, where each behavior $x_i$ contains an item,
timestamp $t_i$, action type, scenario/domain, and side information.
The target ad is $t$ with embedding $\vect v_t\in\R^d$ and category $C(t)\in\{1,\dots,|\mathcal C|\}$.
Let $\vect e_i\in\R^d$ be the semantic embedding of $x_i$.
We define temporal encodings
$\PEpos(i\mid t)\in\R^d$ (relative position to $t$) and
$\PEdel(\Delta t_i)\in\R^d$ for $\Delta t_i = t_t - t_i$.
The temporally encoded behavior is $\tilde{\vect e}_i=\vect e_i+\PEpos(i\mid t)+\PEdel(\Delta t_i)$.
For any vectors $\vect a,\vect b\in\R^d$, $\vect a\odot\vect b$ denotes element-wise product.
Unless stated otherwise, attention $\alpha(\cdot)$ is scaled dot-product:
$\Attn(\vect q,\{\vect k_i\})=\softmax_i\left(\frac{\vect q^\top\vect k_i}{\sqrt d}\right)$.

\subsection{Backbone: Temporal Interest Network (TIN)}\label{subsubsec:backbone}

\begin{figure}
    \centering
    \includegraphics[width=0.98\linewidth]{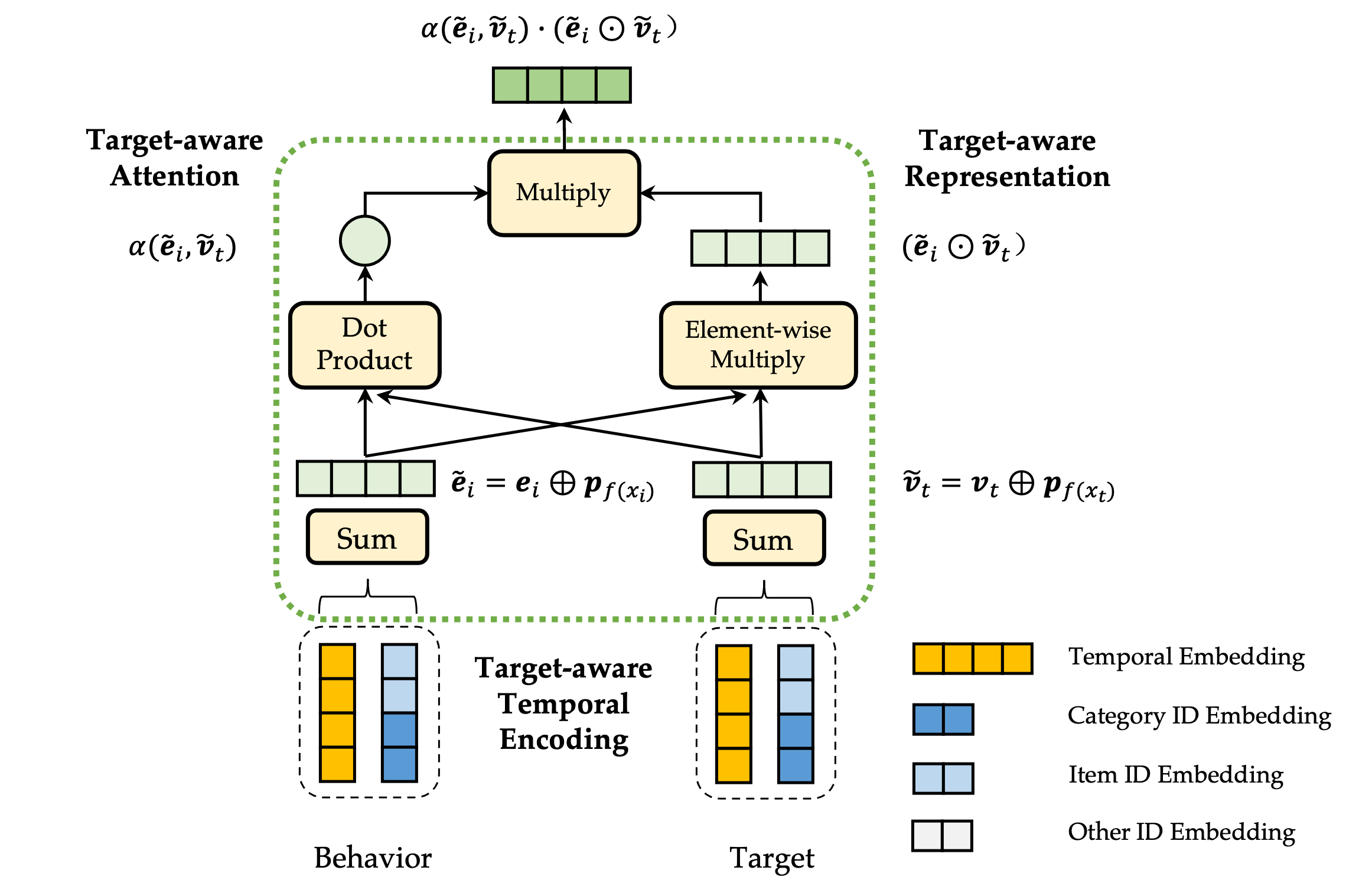}
    \caption{Illustration of the TIN mechanism regarding one behavior. It consists of the target-aware attention, target-aware representation, and target-aware temporal encoding.}
    \label{fig:tin}
\end{figure}

The mainstream user interest models in the ranking are the target-aware attention methods~\cite{zhouguorui2018DIN, zhouguorui2019DIEN, fengyufei2019DSIN}.
In advertisement recommendation, the extreme sparsity of behaviors \emph{makes the temporal information, especially the time interval, more critical} than in the e-commerce or content recommendation.
To this end, we adopt the Temporal Interest Network (TIN)~\cite{zhouhaolin2024TIN} to better capture the temporal correlation between behaviors and the target.
Denote the $\tilde{\vect e}_{i}$ and $\tilde{\vect v}_{t}$ as the semantic-temporal representation of the $i$-th behavior and the target, which will be explained later.
TIN consists of a well-adopted \emph{target-aware attention (TA)} $\alpha(\tilde{\vect v}_{t} \mat W_Q, \tilde{\vect e}_{i} \mat W_K)$ between the $i$-th behavior, and the target $t$ to capture the attentional weight between them, and a novel $d$-dimensional  \emph{target-aware representation (TR)} $(\tilde{\vect v}_{t} \mat W_U \odot \tilde{\vect e}_{i} \mat W_V ) \in \mathcal{R}^d$ as a representation of their interaction.
Formally, the output of the TIN is denoted as:

\begin{align}
    \vect u_{\text{TIN}} & = \sum \alpha(Q, K) \cdot (U \odot V) \\ 
 & = \sum_{x_i \in \mathcal{H}}  \underbrace{\alpha(\tilde{\vect v}_{t} \mat W_Q, \tilde{\vect e}_{i} \mat W_K)}_\text{TA} \cdot \underbrace{(\tilde{\vect v}_{t} \mat W_U \odot \tilde{\vect e}_{i} \mat W_V )}_\text{TR}
 \label{eq:tin}
\end{align}

where  $\mat W_Q, \mat W_K, \mat W_U, \mat W_V \in \mathcal{R}^{d\times d}$ are the projection matrices.
The TA aims to capture the behavior-target correlation, while the TR greatly \emph{improves the discriminability of the representation}~\cite{zhouhaolin2024TIN, feng2024DARE}, compared to the target-agnostic one.
TIN's formulation shares similarities with that of GLU~\cite{shazeer2020glu} and HSTU~\cite{zhai2024hstu} in the sense that \emph{all of them emphasize the element-wise multiplication outside the attention function}.
However, they have the following critical difference: \emph{$Q, U, K, V$ are symmetric in GLU and HSTU since they mainly focus on self-attention, but in our target-attention approach, $Q, U$ correspond to the target, while $K, V$ correspond to the behaviors}.
Refer to Fig.~\ref{fig:tin} for an illustration.

\begin{figure}
    \centering
    \subfigure[Relative Position Pattern]{\includegraphics[width=0.49\linewidth]{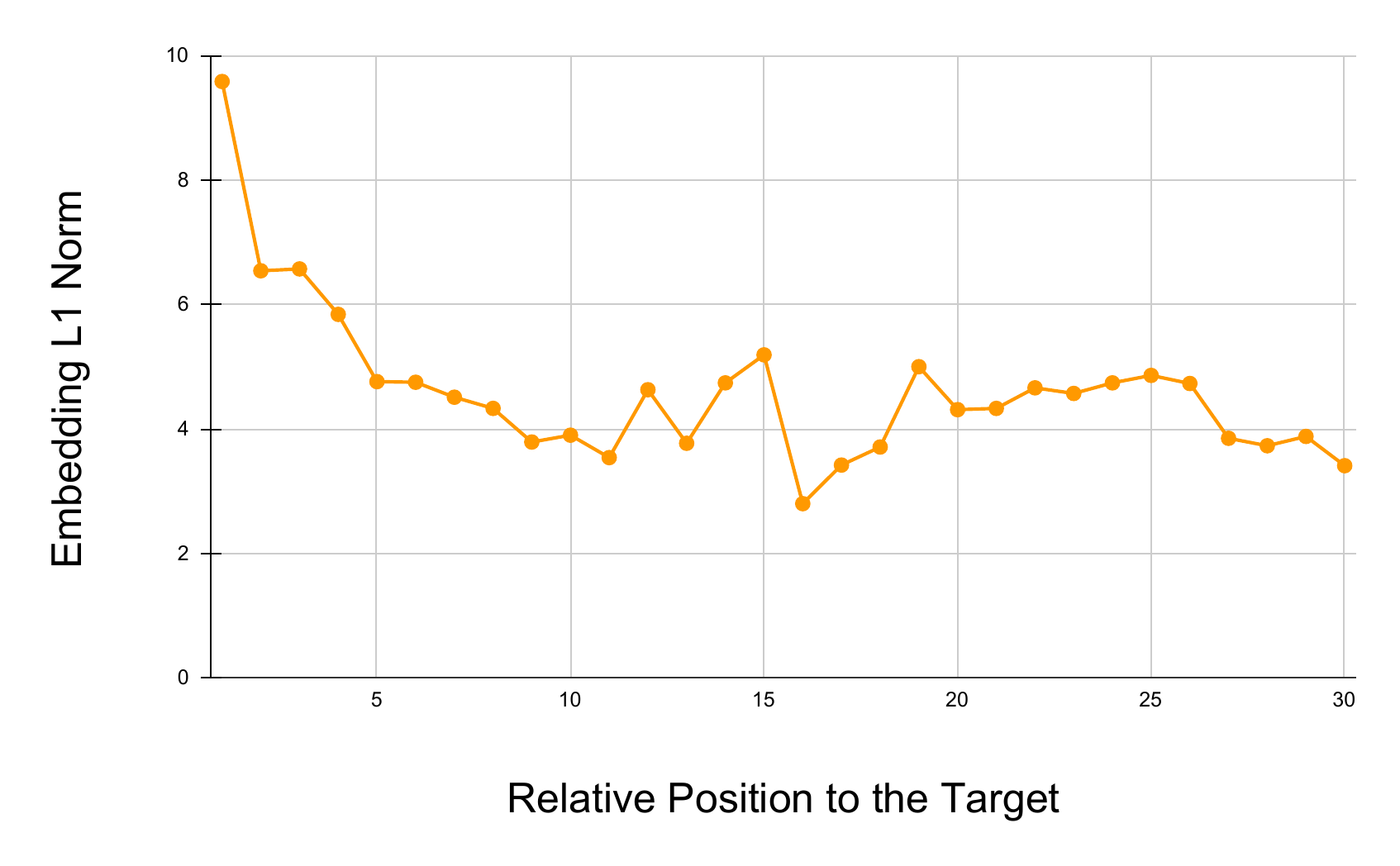}}
    \subfigure[Time Interval Pattern]{\includegraphics[width=0.49\linewidth]{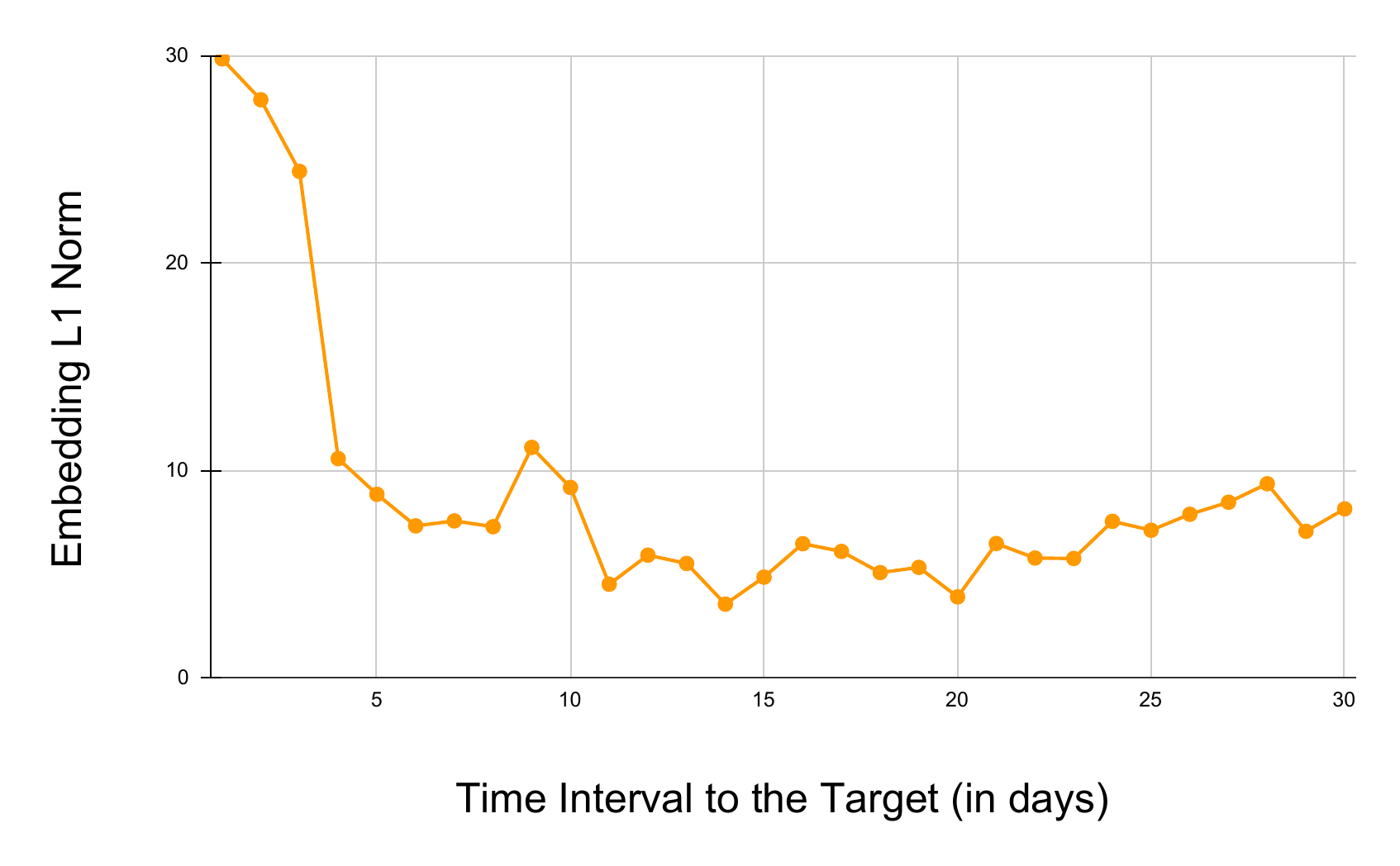}}
    \caption{Learned temporal patterns of our production model. The x-axis denotes each target-aware position (left) or time interval in days between each behavior and the target (right), the y-axis denotes the L1 norm of the corresponding temporal embedding. There is stronger decaying pattern in the time interval embeddings.}
    \label{fig:temporal_pattern}
\end{figure}

We observe that \emph{there are stronger temporal decaying patterns regarding the time interval than the relative position}, as shown in Fig.~\ref{fig:temporal_pattern}.
Employing TIN brings significant GMV lift (\eg, 1.96\% on WeChat Moment~\cite{zhouhaolin2024TIN}) in our primary scenario and was fully launched as the production model for most advertising scenarios since 2023.
All the following model evolutions in ranking are based on the TIN backbone.

\subsection{The Search Policy}\label{subsec:search_policy}

In the following, we present the hierarchical hard search policy for the ranking phase to handle the heterogeneous side features of the sequence, the decoupled embeddings method for soft search, and the stratified sampling method to handle long sequences for matching\footnote{In matching, the candidate set is target‑agnostic; we therefore use category‑wise stratified sampling (not target‑conditioned search). We still place it in the Search section for system symmetry}.

\subsubsection{Hierarchical Hard Search in Ranking}
Existing works mainly rely on one category feature for the hard search~\cite{piqi2020sim}.
In our sequence, with fine-grained features such as categories of various levels in the SPU taxonomy, we are able to conduct a hierarchical hard search based on them to retrieve the most informative behaviors of sufficient length.
Specifically, we first retrieve behaviors with the same 3rd-level (the most fine-grained) category as the target, and if there are not enough results, we then roll back to 2nd and 1st-level categories.
Second, we search the behaviors based on their action types, \emph{e.g.}, impressions, clicks, or conversions, or based on scenarios, \emph{e.g.}, Moments, Channels, or Content, and take a union of the search results.
Finally, if there are still not enough results, we retrieve the latest behaviors since they are, in general, more informative~\cite{zhouhaolin2024TIN}.

\paragraph{Online A/B Testing} 
The hierarchical hard search is the first fully launched long sequence model in our project, and is then widely adopted in many scenarios and tasks.
During the first online A/B tesing in the pCTR prediction of WeChat Moments, hierarchical hard search outperformed the naive category-wise hard search (\emph{i.e.}, SIM hard search) with a relative AUC lift of 0.05\% and GMV lift of 1.73\%.

\subsubsection{Decoupled Embeddings for Soft Search in Ranking}
Existing soft search methods~\cite{piqi2020sim, si2024TWINV2} mainly rely on an attention mechanism to model the importance of each behavior and to retrieve the most informative ones to feed into the downstream user interest model.
However, we identify the conflicts between the attention $\alpha(\vect e_i^S,\vect v_t^S)$ and the representation module $\vect e_i^S \odot \vect v_t^S$ when they share the embeddings~\cite{feng2024DARE}, that is, all embeddings are from the same space: $\vect e_i^S, \vect v_t^S \in \mathcal E^S$.
To this end, we propose a Decoupled Attention and Representation Embeddings (DARE) model~\cite{feng2024DARE}, which decouples the attention and representation module by using two separate embedding tables.
Specifically, we employ two embedding spaces, one for the attention, \emph{i.e.}, $\mathcal E^A$, and one for the representation,  \emph{i.e.}, $\mathcal E^R$.

\begin{align}
    \vect u_{\text{DARE}} & = \sum \alpha(Q, K) \cdot (U \odot V) \\ 
 & = \sum_{X_i \in \mathcal{H}}  \underbrace{\alpha(\tilde{\vect v}_{t}^A \mat W_Q, \tilde{\vect e}_{i}^A \mat W_K)}_{\text{TA in the Att space: }\mathcal E^A} \cdot \underbrace{(\tilde{\vect v}_{t}^R \mat W_U \odot \tilde{\vect e}_{i}^R \mat W_V )}_{\text{TR in the Repr space: }\mathcal E^R}
\label{eq:tin}
\end{align}

In the search phase, we only conduct soft search upon the similarities in the attention embedding space, i.e., $\alpha(\tilde{\vect v}_{t}^A \mat W_Q, \tilde{\vect e}_{i}^A \mat W_K)$.

\paragraph{Online A/B Testing} During the online validation phase, DARE achieved a 1.47\% GMV lift in the pCTR prediction of WeChat Moments, compared with the TWIN baseline, which uses a shared embedding for both attention and representation.

\subsubsection{Stratified Sampling in Matching}
In the matching stage, with a much larger candidate pool, it's impossible to conduct retrieval based on each candidate within the given latency limit.
Therefore, we resort to a target-agnostic strategy, \emph{i.e.}, we utilize the \emph{Category-wise Stratified Sampling} method to sample from the user's ultra-long sequence according to the product category of each behavior, ensuring that behaviors of each category are sampled. 
We chose the category as the stratification feature since it's one of the most important features.

\paragraph{Online A/B Testing} This approach resulted in a 1.5\% GMV lift in the matching of WeChat Channels, and 0.80\% GMV lift of WeChat Moments, compared to a latest-K retrieval baseline.

\subsection{Sequence Modeling}\label{subsec:sequence_modeling}

In this section, we'll first present the Decoupled Side Info TIN and Target decoupled Position Encoding for matching, as well as target decoupled SASRec for matching.
Lastly, we'll demonstrate Stacked TIN to capture high-order interaction.

\subsubsection{Decoupled Side Info TIN (DSI-TIN) for Ranking}\label{subsubsec:DSI-TIN}

\begin{figure}[h!]
    \centering
    \includegraphics[width=0.95\linewidth]{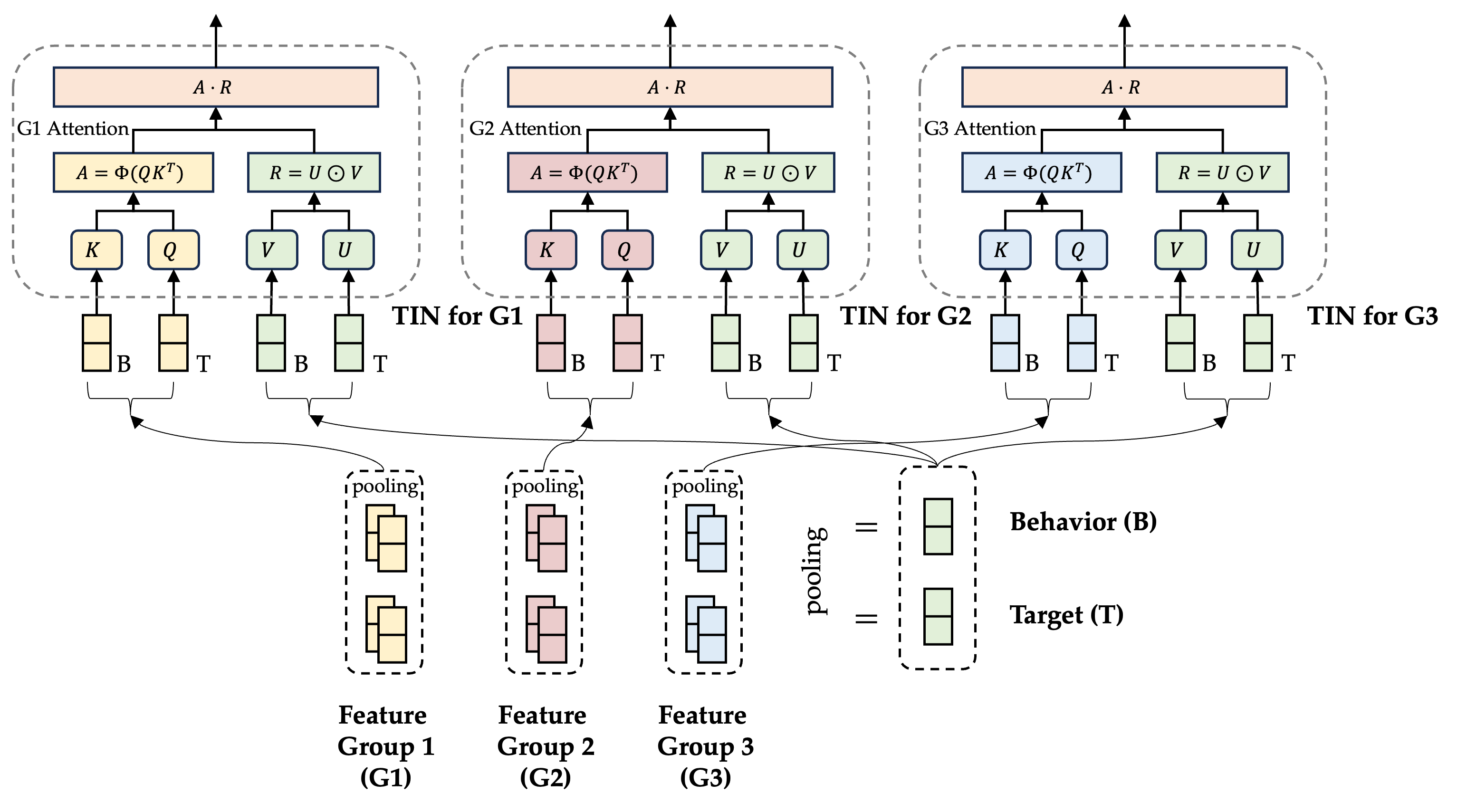}
    \caption{Architecture of Decoupled Side Info TIN. We build multiple TINs, each with a different set of features in the attention module. All TINs use all the side features in their representation.}
    \label{fig:decoupled_side_info_tin}
\end{figure}

When modeling the retrieved sequence after the search stage, we found that due to the diversity of behavior sources, side info of behaviors contains heterogeneous feature fields such as temporal information, categories, and behavior action types. 
Fusing all these feature domains together to calculate Target-Attention will introduce inter-field interference~\cite{tupe2020rethinking, liu2021NOVA, xie2022DIF-SR}.

To this end, inspired by \cite{xie2022DIF-SR, liu2021NOVA, lin2024MSSR, tupe2020rethinking}, we employ multiple TINs, each of which only utilizes a subset of side info as the $Q$ and $K$ in the attention function, while all TINs fuse all side info in their representation.
We name this method Decoupled Side Info TIN (DSI-TIN).
The split strategy among multiple TINs groups together semantically similar side info to the same TIN, while separating those that are different.
Refer to Fig~\ref{fig:decoupled_side_info_tin} for an illustration of DSI-TIN with three subsets \texttt{G1, G2}, and \texttt{G3}.

\begin{table*}[htbp]
\caption{Decoupled Side Info TIN offline evaluation. Report the mean and standard deviation(in brackets) of the results.}
\begin{tabular}{c|cccc|cccc}
\hline
                & \multicolumn{4}{c|}{Alipay}                                                                                                       & \multicolumn{4}{c}{Taobao}                                                                                                       \\ 
\midrule
model           & \multicolumn{1}{c|}{AUC}                     & \multicolumn{1}{c|}{$\delta$\%}   & \multicolumn{1}{c|}{LogLoss}                 & $\delta$\%    & \multicolumn{1}{c|}{AUC}                     & \multicolumn{1}{c|}{$\delta$\%}   & \multicolumn{1}{c|}{LogLoss}                 & $\delta$\%    \\ 
DIF             & \multicolumn{1}{c|}{0.8258(0.0019)} & \multicolumn{1}{c|}{-7.54} & \multicolumn{1}{c|}{0.9779(0.0031)} & +28.69 & \multicolumn{1}{c|}{0.8668(0.0012)} & \multicolumn{1}{c|}{-6.42} & \multicolumn{1}{c|}{0.8598(0.0065)} & +40.67 \\ 
NOVA            & \multicolumn{1}{c|}{0.8352(0.0031)} & \multicolumn{1}{c|}{-6.48} & \multicolumn{1}{c|}{0.9140(0.0106)} & +20.28 & \multicolumn{1}{c|}{0.8627(0.0018)} & \multicolumn{1}{c|}{-6.87} & \multicolumn{1}{c|}{0.8034(0.0065)} & +31.45 \\ 
TIN             & \multicolumn{1}{c|}{0.8931(0.0013)} & \multicolumn{1}{c|}{-}     & \multicolumn{1}{c|}{0.7599(0.0089)} & -      & \multicolumn{1}{c|}{0.9263(0.0003)} & \multicolumn{1}{c|}{-}     & \multicolumn{1}{c|}{0.6112(0.0047)} & -      \\ 
TIN+DSI\_sum    & \multicolumn{1}{c|}{0.8964(0.0021)} & \multicolumn{1}{c|}{+0.37} & \multicolumn{1}{c|}{0.7467(0.0118)} & -1.74  & \multicolumn{1}{c|}{0.9284(0.0005)} & \multicolumn{1}{c|}{+0.23} & \multicolumn{1}{c|}{0.6000(0.0024)} & -1.83  \\ 
TIN+DSI\_concat & \multicolumn{1}{c|}{\textbf{0.9029(0.0011)}} & \multicolumn{1}{c|}{+1.10} & \multicolumn{1}{c|}{\textbf{0.7293(0.0154)}} & -4.03  & \multicolumn{1}{c|}{\textbf{0.9309(0.0003)}} & \multicolumn{1}{c|}{+0.50} & \multicolumn{1}{c|}{\textbf{0.5887(0.0051)}} & -3.68  \\ 
\bottomrule
\end{tabular}
\label{tab:table1}
\end{table*}

\paragraph{Offline Evaluation}

We conduct offline experiments on public datasets, Alipay and Taobao, to validate the effectiveness of our proposed method. 
We compare our proposed Decoupled Side Info TIN with the following baselines: the vanilla TIN without any decoupling, the Nova~\cite{liu2021NOVA} and DIF-SR~\cite{xie2022DIF-SR}. 
Our method (TIN+DSI\_concat) outperforms the best-performing baseline, \ie, the vanilla TIN, with a relative AUC lift of 1.1\% and a relative LogLoss decrease of 4.03\% on the Alipay dataset, and with a relative AUC lift of 0.5\% and a relative LogLoss decrease of 3.68\% on Taobao dataset, as presented in Tab.~\ref{tab:table1}. All these improvements are statistically significant.

\paragraph{Online A/B Testing} In WeChat Channels pCTR prediction, we split features into three groups: the 1st consists of only temporal info; the 2nd consists of the action type, and the last consists of all remaining features: ad ID, category ID, \emph{etc.} The DSI-TIN achieves a 0.58\% GMV lift over a DIF-SR baseline during online A/B test.

\subsubsection{Target Decoupled Temporal Encoding for Ranking}

The TIN employs a global temporal encoding for behaviors.
However, in conversion prediction, we aim to predict users' various types of conversions, ranging from download and order to the final payment.
There may be different temporal patterns of behavior with respect to different types of target conversions.
Therefore, the unified temporal encoding of TIN, \emph{i.e.}, $ \PEpos(i\mid t), \PEdel(\Delta t_i)$, may lead to target-wise interference of these patterns, making it fail to capture such target-specific temporal patterns.
To this end, we propose to disentangle the temporal encoding for different target side info, \textit{e.g.}, the target categories.
Specifically, we assign independent target positions, \textit{i.e.}, position 0, \emph{i.e.}, $\vect q^{\text{pos}}_{C(t)}, \vect q^{\Delta}_{C(t)}\in\R^d$,  for each target type $C(t)$:

\begin{align}
\tilde{\vect e}_i &= \vect e_i + \PEpos(i\mid t) + \PEdel(\Delta t_i), \\
\tilde{\vect v}_t &= \vect v_t + \vect q^{\text{pos}}_{C(t)} + \vect q^{\Delta}_{C(t)},
\end{align}
where $\vect q^{\text{pos}}_{c},\vect q^{\Delta}_{c}\in\R^d$ are learned per category.

\begin{table*}[htbp]
\caption{Decoupled Temporal Encoding offline evaluation. Report the mean and standard deviation(in brackets) of the results.}
\begin{tabular}{c|cccc|cccc}
\hline
         & \multicolumn{4}{c|}{Alipay}                                                                                                      & \multicolumn{4}{c}{Taobao}                                                                                                      \\ \hline
model    & \multicolumn{1}{c|}{AUC}                     & \multicolumn{1}{c|}{$\delta$\%}   & \multicolumn{1}{c|}{LogLoss}                 & $\delta$\%   & \multicolumn{1}{c|}{AUC}                     & \multicolumn{1}{c|}{$\delta$\%}   & \multicolumn{1}{c|}{LogLoss}                 & $\delta$\%   \\ 
TIN      & \multicolumn{1}{c|}{0.8931(0.0013)} & \multicolumn{1}{c|}{-}     & \multicolumn{1}{c|}{0.7599(0.0089)} & -     & \multicolumn{1}{c|}{0.9263(0.0003)} & \multicolumn{1}{c|}{-}     & \multicolumn{1}{c|}{0.6112(0.0047)} & -     \\ 
TIN+TdPE & \multicolumn{1}{c|}{0.8968(0.0008)} & \multicolumn{1}{c|}{+0.41} & \multicolumn{1}{c|}{0.7575(0.0137)} & -0.32 & \multicolumn{1}{c|}{0.9278(0.0003)} & \multicolumn{1}{c|}{+0.16} & \multicolumn{1}{c|}{0.6060(0.0038)} & -0.85 \\ 
TIN+dPE  & \multicolumn{1}{c|}{\textbf{0.8974(0.0005)}} & \multicolumn{1}{c|}{+0.48} & \multicolumn{1}{c|}{\textbf{0.7413(0.0110)}} & -2.45 & \multicolumn{1}{c|}{\textbf{0.9284(0.0002)}} & \multicolumn{1}{c|}{+0.23} & \multicolumn{1}{c|}{\textbf{0.6029(0.0045)}} & -1.36 \\ 
\hline
\end{tabular}
\label{tab:table2}
\end{table*}

\paragraph{Offline Evaluation}
We conduct offline experiments on public datasets to validate the effectiveness of our proposed method. 
We compare our proposed decoupled Temporal Encoding method with the vanilla TIN baseline with one universal temporal encoding. 
Our method(TIN+dPE) gets a relative AUC lift of 0.48\% and a relative LogLoss reduction of 2.45\% on Alipay dataset,  and a relative AUC lift of 0.23\% and a relative LogLoss reduction of 1.36\% on Taobao dataset, as presented in Tab.~\ref{tab:table2}. All these improvements are statistically significant. Since TIN+dPE learns multiple sets of timing information of target and behavior together, its performance is slightly better than TIN+TdPE, but TIN+TdPE has fewer learnable parameters, so it is also competitive.

\paragraph{Online A/B Testing} We mainly decouple temporal embeddings for the conversion prediction model, since the correlation varies a lot for conversions with different categories.
During online test in pCVR prediction in WeChat Moments, this method achieves a 0.79\% GMV lift, compared to the TIN baseline without decoupling..

\subsubsection{Target Decoupled SASRec for Matching}\label{subsubsec:target_SASRec}

In matching, for the sake of optimizing performance, the two-tower structure is usually used, that is, constructing the user tower and the advertising tower separately. 
However, this two-tower structure fails to model users' multiple interests regarding the target, \ie,  which is critical in long-sequence modeling.
Many work~\cite{li2019MIND, cen2020ComiRec, tan2021SINE, xu2022MKVE, zhai2023revisiting, ding2025retrieval} have been proposed to tackle this challenge to extract multiple user interests, but these models are target-agnostic.

We propose a novel user interest modeling method for matching upon the sampled sequence via category-wise stratified sampling to tackle the target-wise interference challenge of the vanilla two-tower architecture.
In particular, we capture the correlations between behaviors by SASRec~\cite{kangwangcheng2018SASRec}, 
then employ $|\{C_i\}|$ output heads, each corresponding to a specific ad category.
Note that $|\{C_i\}|$ is around 50 in our system.
Only the user head corresponding to the target ad's category is activated and multiplied by the ad embedding from the ad tower to get the final prediction score.
We name this method Target Decoupled SASRec since we decouple the output of user towers by target categories. 
Formally,

\begin{equation}
    \Phi = \langle \bm{f}^{C(t)}_\text{SASRec}(\{\vect e_i\}),  \vect v_t \rangle
\end{equation}

where $C(t)$ denotes the category of target $t$, and $\bm{f}^{C(t)}_\text{SASRec}(\{\vect e_i\})$ denotes the $C(t)$-th output vector of the target decouple SASRec.

\begin{figure}[h!]
    \centering
    \includegraphics[width=0.95\linewidth]{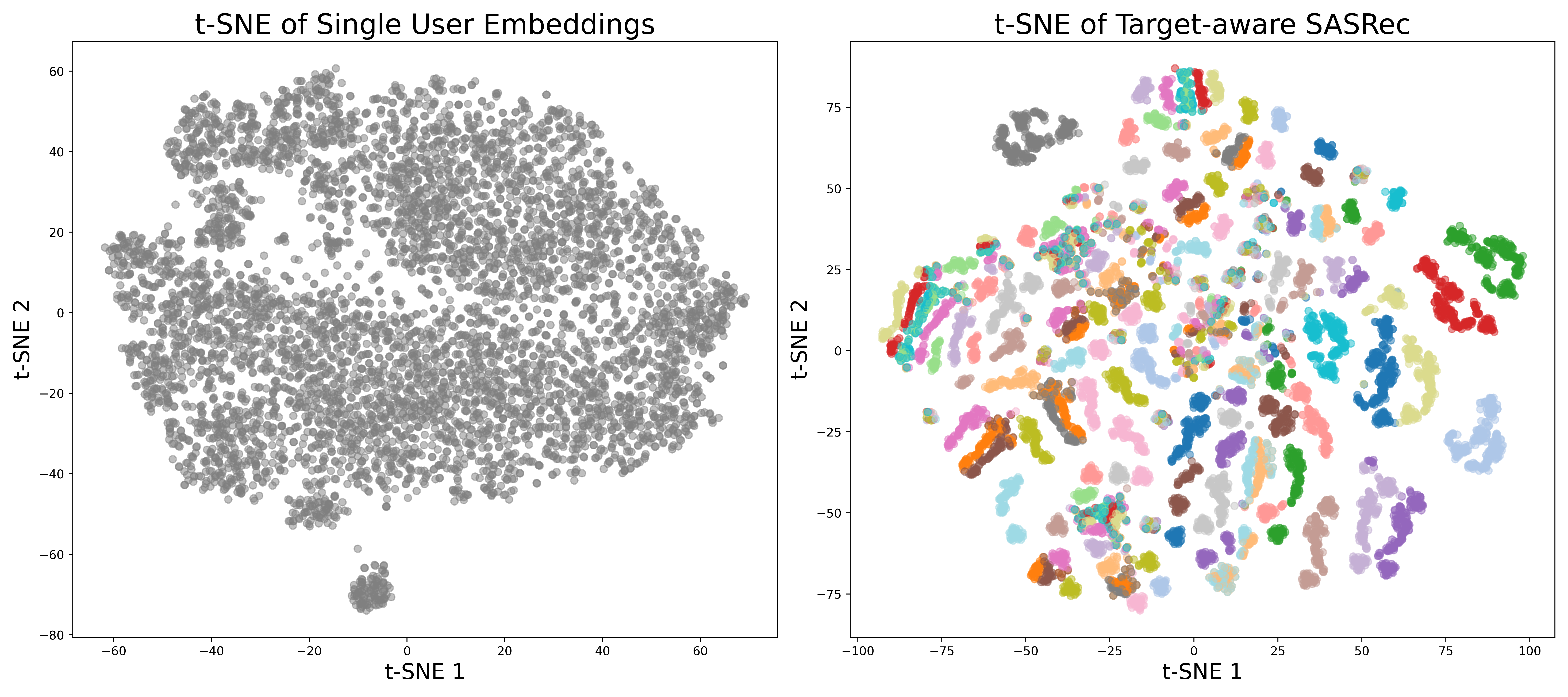}
    \caption{Visualization of the outputs from the vanilla SASRec user tower (left) and the Target decoupled SASRec tower (right) through t-SNE. Colors denote ad categories. There is a better clustering pattern in our Target-aware SASRec.}
    \label{fig:pre_ranking_cluster}
\end{figure}

We compared the output of the traditional SASRec and our Target decoupled SASRec. Specifically, we reduced the dimensions of these embeddings through the t-SNE and colored them according to the category, as shown in Fig.~\ref{fig:pre_ranking_cluster}. 
It can be clearly seen that the outputs learned by the vanilla SASRec method lack cluster information, while our method can obviously learn structured clusters, and these clusters have good correspondence with the advertising categories: all points of some clusters belong to the same advertising category.

\paragraph{Online A/B Testing} This method contributed a 0.70\% GMV lift in the matching of WeChat Channels, with 39 target categories plus one default category.

\subsubsection{Stacked TIN (STIN) for High-order Interaction Ranking}\label{subsubsec:stacked_TIM}

In our system, the user's historical behavior sources are complex, and it's necessary to capture high-order interactions based on the type, scenario, domain, and other fine-grained side info, such as \texttt{(Behavior Category X Target Category X Behavior Scenario X Target Scenario)}.
Therefore, we aim to build a multi-layer sequential recommendation model to capture high-order interactions between behaviors and the target.

Some works~\cite{kangwangcheng2018SASRec} already achieve this for matching~\cite{kangwangcheng2018SASRec} by stacking the self-attention module~\cite{vaswani2017Transformer}, but it's still an open research area in the ranking.
There are some attempts~\cite{chenqiwei2019BST, zhai2024hstu} to append the target to the end of the behavior sequence, and then perform self-attention upon this new sequence.
However, the performance of this stacking paradigm is not as efficient as the target-attention methods~\cite{zhouhaolin2024TIN}.
On the other side, existing target-attention methods~\cite{zhouguorui2018DIN, zhouguorui2019DIEN, fengyufei2019DSIN} rely on the MLP to model the relationships between behaviors and the target in the representation layer.
But ~\cite{rendle2020NCF_revisited} has proved that MLP is hard to learn the dot product.

To this end, we propose a \emph{novel stacking strategy to construct multi-layer target attention models}.
Unlike self-attention, where the inputs $ Q$, $ K$, and $ V$ are symmetric, in target attention, the inputs have distinct roles.
Taking TIN (Eq.~\ref{eq:tin}) as an example, the $K, V$ correspond to the behavior, while $Q, U$ correspond to the target. 
When stacking target-attention, we also separate the inputs as well as the outputs between behavior and target.

\begin{figure}
    \centering
    \includegraphics[width=0.95\linewidth]{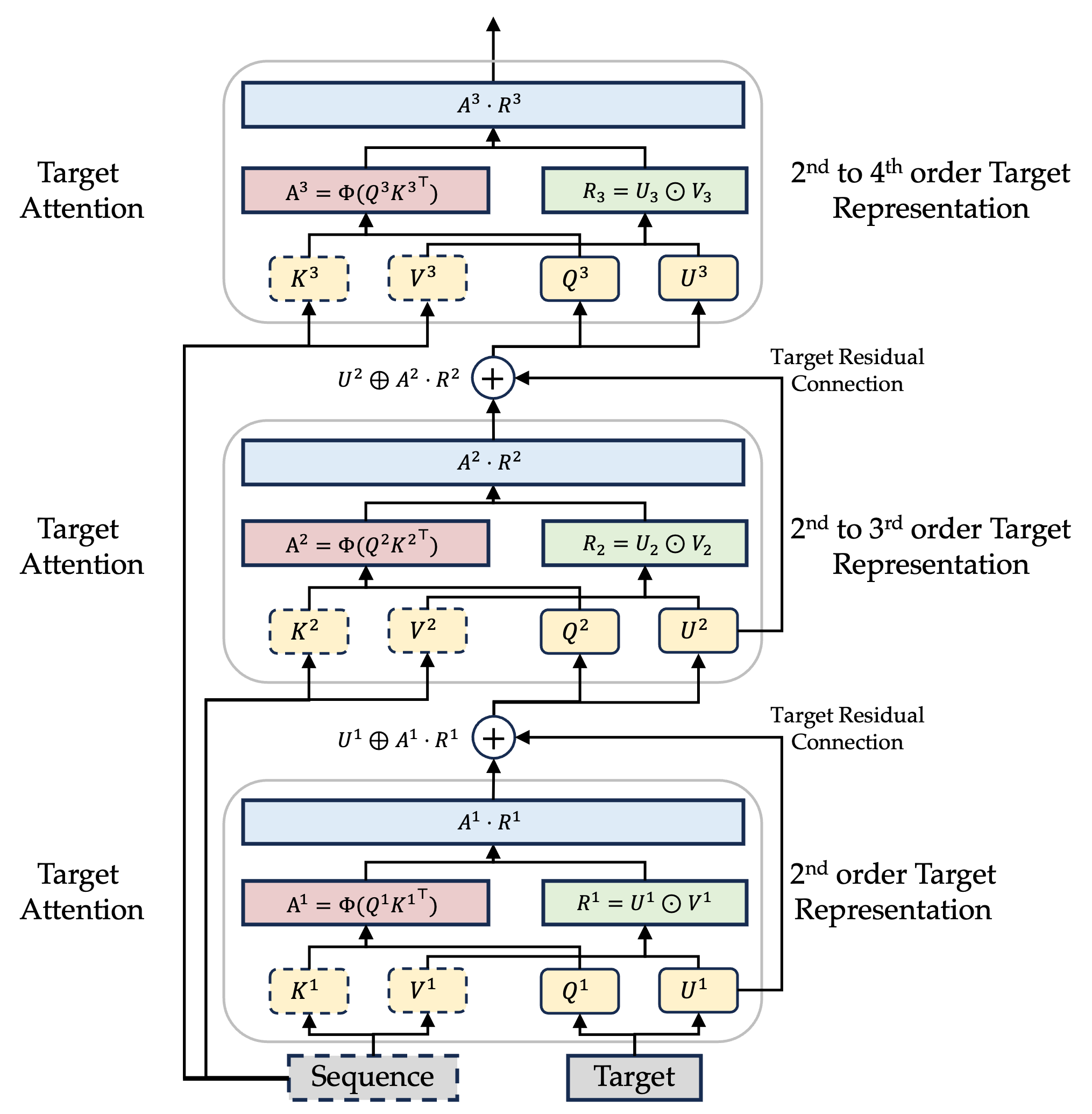}
    \caption{Architecture of Stacked TIN. We utilize the Temporal Interest Network as the stacking block. In the $l$-th layer, the $K$ and $V$ correspond to the behaviors, and $Q$ and $U$ correspond to the output of the previous layer, with a skip connection to the target.}
    \label{fig:stacked_TIM}
\end{figure}

In the $l$-th layer, we define the $K$ and $V$ as the transformed representation of user behaviors $\vect e_i \in \mathcal{R}^d$, through projection matrix $\mat W_K^l, \mat W_V^l \in \mathcal{R}^{d \times d}$.
The $Q$ and $U$ denote the target: in the first layer, they are simply the target $\tilde{\vect v}_{t}$ projected by matices $\mat W_Q^1, \mat W_U^1 \in \mathcal{R}^{d \times d}$; while in the $l$-th layer, they are defined as the output of the previous layer, \ie, $\vect u_{\text{TIN}}^{l-1}$, with a residual connection to the target $\tilde{\vect v}_{t}$. 
Formally, the output of the $l$-th layer is :

\begin{align}
    \vect u_{\text{TIN}}^l &= \sum \alpha(Q^l, K^l) \cdot (U^l\odot V^l)\\
    Q^l &= (U^{l-1} \oplus \vect u_{\text{TIN}}^{l-1}) \mat W_Q^l \\ 
    U^l &= (U^{l-1} \oplus \vect u_{\text{TIN}}^{l-1}) \mat W_U^l \\
    K^l &= \tilde{\vect e}_{i} \mat W_K^l \\ 
    V^l &= \tilde{\vect e}_{i} \mat W_V^l
\end{align}

where $\mat W_K^l, \mat W_V^l, \mat W_Q^l, \mat W_U^l \in \mathcal{R}^{d \times d}$ denote the projection matrix, $\oplus$ denotes element-wise addition; $Q^1 = \tilde{\vect v}_{t} \mat W_Q^1, U^1 = \tilde{\vect v}_{t} \mat W_U^1 $. With the residual connection, the representation of the $l$-th layer captures 2nd to $l+1$ order cross interactions between target and behaviors.
Refer to Fig.~\ref{fig:stacked_TIM} for an illustration.

\paragraph{Offline Evaluation}
We conducted offline experiments on the Amazon dataset\footnote{http://jmcauley.ucsd.edu/data/amazon/}. 
Our findings demonstrate that stacking 2 and 3 TINs using the proposed method resulted in an improvement in AUC from 0.8736 to 0.8767 and 0.8778, respectively, corresponding to a relative lift of 0.355\% and 0.125\%. 
Interestingly, further stacking of TINs did not yield significant performance gains, suggesting that the behavior patterns in this dataset are relatively simple. 
Specifically, a 3-layer TIN proved to be sufficient in capturing these patterns, while a 4-layer TIN suffered from overfitting. 

\paragraph{Online A/B Testing}
We design a 4-layer STIN , which achieves a 1.01\% GMV lift in the pCTR prediction of WeChat Channels, and 0.65\% GMV lift in the pCTR prediction of WeChat Moments, compared to the single-layer TIN production model.
We analyze the learned attention weight in different layers, and do observe different attention patterns across layers: the weight distribution gradually becomes polarized, with an increasing proportion of weights become very large or small.

\section{Platform}
\label{sec:platform}

In this section, we will describe the system optimization for both sparse parameters and dense computation to support the transformation of the advertising system from short-sequence modeling with hundreds of behaviors to full-domain, full-historical long-sequence modeling at the level of tens of thousands.

\begin{figure}
    \centering
    \includegraphics[width=0.95\linewidth]{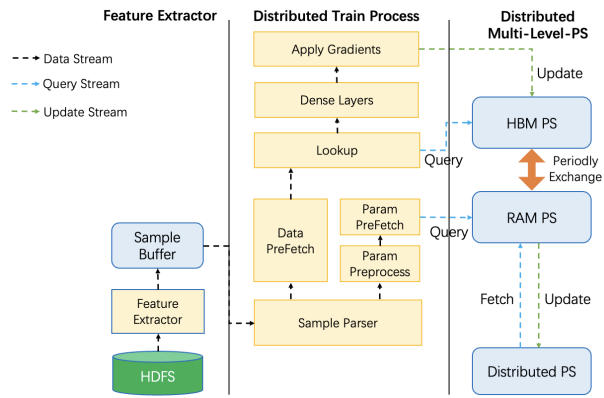}
    \caption{The overview of modules during model training.}
    \label{fig:platform_overview}
\end{figure}

\subsection{Overview}

For the sparse parameters such as ID embeddings, we first parse the samples and their features through a Sample Parser and then prefetch the data and parameters into the RAM and HBM.
During the forward computation, we optimize the parameter lookup and GPU kernels by Shared-Memory Access and GPU Multi-Stream.
Then, for the dense computation, we focus on accelerating the memory-access-intensive operators via our Heterogeneous Engine Acceleration.
Refer to Fig.~\ref{fig:platform_overview} for an overview of the platform.

\subsection{System Optimization for Sparse Parameters}


GPU Acceleration typically involves three main steps: Firstly, the data and model parameters are prefetched from the dataset and then copied into the GPU memory. 
Secondly, the GPU executes the forward/backward computations of the model, including the lookup of sparse parameters; 
Finally, the result is transferred from the GPU memory to the system memory. In this section, we will describe our customized optimization strategies in the above steps when tailoring long sequence modeling into industrial online models.

\subsubsection{Parameter and Data Prefetch}

\paragraph{Parameter Preprocess}
An important task before parameter prefetch is to preprocess the parameters, which involves the deduplication of sparse features and the calculation of frequency statistics. 
This part of the work consumes a lot of CPU resources. With the sequence becoming longer, the number of features per sample increases by 3 to 5 times. 
We found that there is a significant bottleneck in CPU load, up to 85\%, causing the sample processing performance to come down by a factor of $10$. Particularly, the sub-module of key collection for feature de-duplication and frequency statistics dropped the most, so we upgraded it from CPU-based multi-threading Parallel-Hash-Map to GPU High-Performance Accelerating Operations

\paragraph{Parameter Prefetch}

With the increase of sequence length, the number of sparse ID features also increases greatly, increasing the number of sparse parameters, \emph{i.e.}, embeddings. 
Therefore, the prefetching of sparse parameters puts greater pressure on the system. 
Specifically, a longer parameter exchange time between multi-level caches, \emph{i.e.}, High-Bandwidth Memory (HBM), RAM, and Disk, causes a certain degree of performance bottleneck. 
After analyzing real data in some typical scenarios, we found that the feature overlap rate in different scenarios ranges from $60\%$ to $75\%$, which is relatively high.  

The commonly used mechanism of full parameter swap-in and swap-out has resulted in a significant amount of bandwidth waste. 
We introduce a new mechanism called \emph{Incremental-Exchange} to incrementally exchange embedding parameters between RAM and HBM.
Specifically, 
it introduces three designs in the pipeline of the training process.
\emph{1) Incremental swap-in}: Incremental Embedding parameters are swapped from CPU memory to GPU memory;
\emph{2) LRU swap-out}: Based on the amount of memory used, we select the embedding parameters to be swapped out from GPU memory to CPU memory in the Least Recently Used (LRU) manner;
\emph{3) High-frequency Embedding sustain
}: High-frequency Embedding parameters are sustainably held in the GPU memory;

\paragraph{Data Prefetch}
In addition, we found that there is another significant bottleneck in {Data-Prefetch}, and the time for copying samples from Host to Device has increased by three times, in which the copy time accounts for a high proportion. 
We optimized it in two steps: First, we introduced \emph{pinned memory} and upgraded synchronous copy to \emph{asynchronous copy}. 
Secondly, we arranged loading, parsing, and pre-processing in a pipeline to overlap the asynchronous copy time

\subsubsection{Forward Computation}

\paragraph{Parameter Lookup}
In the sparse parameter lookup module, as the number of sparse ID features increases, the query amount of sparse embedding parameters rises thrice, causing a chain bottleneck. 
Meanwhile, as the latency of forward lookup and backward gradient merging increases, the utilization of Memory increases by more than $85\%$, resulting in a typical memory bound. 
It is well known that the access speed of Shared Memory is 100 times faster than that of Global Memory. 
For multiple global memory accesses with the same address, we replace \emph{Multiple Global Memory Access} with \emph{Single Global Memory Access \& Multiple Shared-Memory Access}.

\paragraph{Multi-Stream}
Next, as the sequence becomes longer, the number of small kernels in the lookup operator increases significantly, \textit{e.g.}, the forward combiner kernel and the backward gradient index generation kernel, which doubles time consumption. 
Therefore, we introduced a \emph{Multi-Stream} optimization, which increases the computational parallelism by launching a large number of small Kernels to different streams for parallel execution, thereby improving the utilization of the Stream-Multi processor. 
Secondly, we used CudaGraph optimization, upgrading "multiple separate Kernel Launches" to "one overall Kernel Launch", significantly improving the launch bottleneck of numerous small Kernels.

\subsection{System Optimization for Dense Computation}

\begin{figure*}
    \centering
    \subfigure[The Latency Distribution of Dense Model.]{\includegraphics[width=0.38\linewidth]{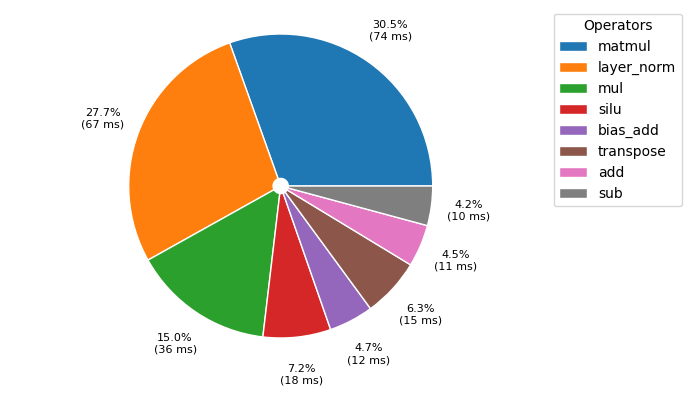}}
    \subfigure[Ablation of Optimization for Training.]{
    \includegraphics[width=0.3\linewidth]{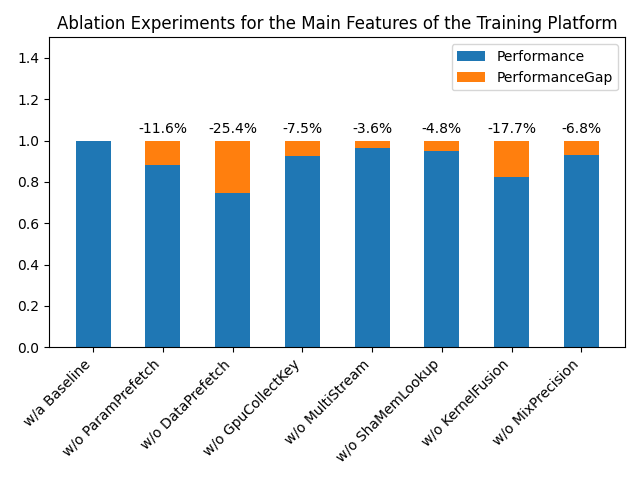}}
    \subfigure[Ablation of Optimization for Inference.]{\includegraphics[width=0.3\linewidth]{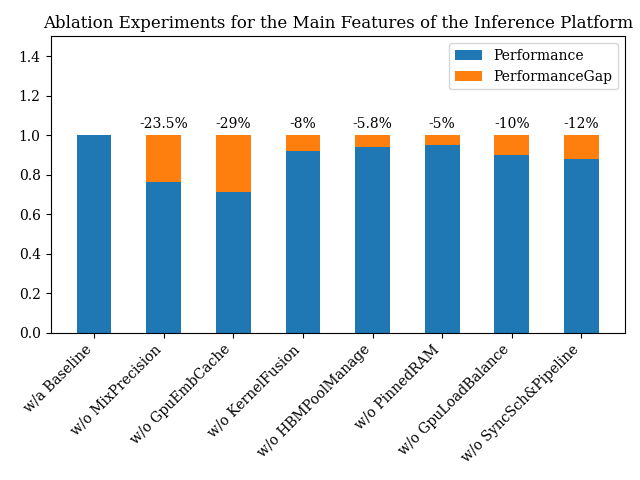}}
    \caption{(a): Only MatMul belongs to computation-intensive OP, the rest of the OPs belong to memory-access intensive OP. 
    (b): In training, Data Prefetch, Kernel Fusion, and Parameter Prefetch are the most important optimizations,  contributing to a total 54.7\% performance improvement. (c): For inference, Mix Precision and GPU Embedding Cache are the most important, contributing to 62.5\% performance improvement.}
    \label{fig:pies_and_bars}
\end{figure*}

\paragraph{Latency Analysis of Dense Computation}
We delve into the cost of different operators and find that the memory-access-intensive ones, which require frequent access to storage units, occupy 70\% of the latency, mainly including element-wise computation. 
While the computation-intensive operators, which require a large amount of floating-point operations, occupy 30\% of the latency, such as Conv, LSTM, FC, MatMul, etc. 
Through the analysis from Fig.~\ref{fig:pies_and_bars}(a), we found that the number of memory-intensive operators has increased, and the time consumption of Element-wise operators: LayerNorm/Mul/Silu/Add has increased by 2.8 times compared to non-sequence models.
We'll present our practices to accelerate these operators in the next paragraph.

\paragraph{Heterogeneous Engine Acceleration}

In our scenario, developers have defined many complex model structures and operations, and the platform can't optimize each one individually. 
We adopt the following solutions to accelerate the train/inference engine.
First, for the general computing acceleration, we used the \emph{mixed precision}, which leverages Tensor Cores to accelerate model computations, achieving a 23\% performance gain. 
Second, for the general memory access optimization, we \emph{replaced the CPU with a GPU to query hot parameters}.
We employed automatic graph computation analysis and further fused memory-access-intensive operators. 
These two techniques improve the inference efficiency by 29\% and 8\%, respectively. 
Third, regarding the heterogeneous resource management, we abandoned static GPU memory allocation and adopted \emph{dynamic GPU memory pool management} to further improve GPU memory utilization, yielding a 5.8\% gain. 
Fourth, for CPU/GPU scheduling, we adopted \emph{asynchronous pipeline scheduling} to reduce blocking and enhance performance by more than 10\%.
The performance of matching and ranking in multiple scenarios has been improved by more than 5 times, and the cost of landing has been greatly reduced.

Considering the complexity of the training and inference platform, a single performance cannot reflect the benefits of features on the entire system. 
We conducted an ablation study on each optimization of the platform to see the impact on the overall performance of the training or inference. 
In model training, the three main optimizations are Data Prefetch, Kernel Fusion, and Parameter Prefetch, which in total contributed to 54.7\% performance improvement. 
In inference, Mix Precision and GPU Embedding Cache contributed to 62.5\% performance improvement
Refer to Fig.~\ref{fig:pies_and_bars}(b) and Fig.~\ref{fig:pies_and_bars}(c)  for details.

Finally, in multiple sequence modeling scenarios of the business, both training and inference have achieved 3 to 5 times performance improvement. In the case of the model complexity increasing by 5 times and the sparse feature increasing by 100 times, the training and inference resources have not increased.

\section{Related Work}\label{sec:related_work}

\paragraph{Sequence Model for Ranking}
Deep Interest Network (DIN)~\citep{zhouguorui2018DIN} introduces target-aware attention, using an MLP to learn attentive weights of each history behavior regarding a specific target. 
This framework has been extended by models like DIEN~\citep{zhouguorui2019DIEN}, DSIN~\citep{fengyufei2019DSIN}, and BST~\citep{chenqiwei2019BST} to capture user interests better. 
~\cite{zhouhaolin2024TIN} proposed to involve the target-aware temporal encoding and representation, significantly improving the model's ability to capture semantic and temporal correlation.

\paragraph{Decouple Side Info}

There are many works to decouple the side info~\cite{RecSys16-pRNN, zhoukun2020s3-rec, SIGIR21-ICAISR, liu2021NOVA, xie2022DIF-SR, IJCAI19-FDSA, IEEE-FDSA-CL}. For example, 
NOVA-SR~\cite{liu2021NOVA} proposes a non-invasive attention mechanism to enhance the attention calculation.  
DIF-SR~\cite{xie2022DIF-SR} decouples the attention calculation of items and side information to attain the fused attention matrices. 
FDSA~\cite{IJCAI19-FDSA} and its enhanced version~\cite{IEEE-FDSA-CL} use separate self-attention layers to extract the representations at two levels, which are then concatenated and fed into the prediction layer.
Similarly, C\footnotesize{A}\normalsize{F}\footnotesize{E}\normalsize~\cite{SIGIR2022-CAFE} takes the sum of two representations to calculate the prediction score.

\paragraph{Matching}

Both MIND~\cite{li2019MIND} and ComiRec~\cite{cen2020ComiRec} aim to capture users' multiple interests based on their behavior sequences, mainly for retrieval.
SINE~\cite{tan2021SINE} ensures training and inference consistency by extracting users' top-K interests and aggregating them. 
MVKE~\cite{xu2022MKVE} addresses the issue of training and inference consistency, but it does not model users' multiple interests based on behavior sequences. 
Additionally, inference requires exhaustive calculations, which increases computational overhead.

\paragraph{Long Sequence Modeling}
SIM~\citep{piqi2020sim} is a pioneering work in this are which introduces the two-stage paradigm.
Models like ETA~\citep{chenqiwei2021ETA} and SDIM \citep{cao2022sampling} further improve this framework. 
Notably, TWIN~\citep{si2024TWINV2} and TWIN-V2~\citep{si2024TWINV2} unify the target-aware attention metrics used in both stages, significantly improving search quality.
Recently, DARE~\cite{feng2024DARE} proposed to decouple the embeddings for the attention and representation module to avoid optimization conflicts.
~\cite{hou2024cross} employed the cross representation production to bridge representations across domains, and the lifelong attention pyramid with three levels of cascading attentions to extract user's interest.
~\cite{chai2025longer} proposed a GPU-efficient long sequence recommender.

\section{Conclusion}

This paper presents our practices for long-sequence modeling in Tencent's Advertising Platform.
We first illustrate how to construct unified commercial behavior trajectories, and then present our modeling and platform innovations.
We succeed in deploying long-sequence advertising models, leading to one of the largest GMV lift in the last decade.





\bibliographystyle{ACM-Reference-Format}
\bibliography{8.reference}


\end{document}